\pdfoutput=1 

\documentclass[aps,twocolumn,prl,showpacs,amsmath,amssymb,nofootinbib,floatfix,nolongbibliography]{revtex4-2}
\usepackage{graphicx,xcolor}
\usepackage{scrextend}
\usepackage{manfnt,pifont}
\usepackage{amsmath}
\usepackage{mathrsfs}
\usepackage{url}
\usepackage{lipsum}
\usepackage{booktabs}
\usepackage[T1]{fontenc}

\definecolor{dark-red}{rgb}{0.6,0,0}
\definecolor{dark-green}{rgb}{0.08,0.36,0.06}
\definecolor{dark-blue}{rgb}{0.3,0.3,0.7}

\usepackage{tikz}

\usepackage{enumitem}
\setlist[itemize]{leftmargin=*}
\usepackage{hyperref}

\hypersetup{
  linktocpage,
  colorlinks  = true, 
  urlcolor    = dark-green, 
  linkcolor   = dark-red, 
  citecolor   = dark-blue 
}

\def\e{{\mathrm e}}
\def\dd{{\mathrm d}}
\def\ds{{\rm ds}}

\def\SU{{\rm SU}}

\newcommand\wh\widehat

\newcommand{\vol}{\mathrm{vol}}
\renewcommand{\Im}{\mathrm{Im}}

\makeatletter 
    
\renewcommand\onecolumngrid{
\do@columngrid{one}{\@ne}%
\def\set@footnotewidth{\onecolumngrid}
\def\footnoterule{\kern-6pt\hrule width 1.5in\kern6pt}%
}

\renewcommand\twocolumngrid{
        \def\footnoterule{
        \dimen@\skip\footins\divide\dimen@\thr@@
        \kern-\dimen@\hrule width.5in\kern\dimen@}
        \do@columngrid{mlt}{\tw@}
}%

\makeatother    

\makeatletter
\renewcommand{\section}{%
  \@startsection{section}{1}{\z@}%
    {-3.5ex plus -1ex minus -.2ex}
    {2.3ex plus .2ex}
    {\centering\normalfont\bfseries}
}
\makeatother

\begin{document}
\title{Scale-Covariant Holography}

\author{Nikolay Bobev$^a$}
\author{Pieter Bomans$^b$}
\author{Fri\dh rik Freyr Gautason$^c$}
\affiliation{\vspace{3pt}$^a$Instituut voor Theoretische Fysica and Leuven Gravity Institute, KU Leuven, Celestijnenlaan 200D, B-3001 Leuven, Belgium}
\affiliation{$^b$Deutsches Elektronen-Synchrotron DESY, Notkestr. 85, 22607 Hamburg, Germany}
\affiliation{$^c$STAG Research Centre \& Mathematical Sciences, University of Southampton, Highfield, Southampton SO17 1BJ, U.K.}

\begin{abstract} 
\noindent We explore a vast landscape of scale-covariant string backgrounds that arise from the backreaction of D$p$-branes. In a convenient Weyl frame the metric has an AdS factor and the dilaton has a non-trivial profile. This structure provides a notion of scale covariance that facilitates a holographic analysis and provides a window into the dynamics of the dual strongly coupled non-conformal gauge theory. We present different classes of supergravity solutions of this kind and discuss their dual field theory interpretation. We find several new supersymmetric backgrounds arising from D2-branes by utilizing four-dimensional gauged supergravity, D$p$-branes wrapped on compact manifolds, and infinite families constructed directly in ten dimensions. We construct holographic RG flows between some of the new supersymmetric backgrounds and compute the spectrum of low-lying local operators. Holography determines the form of observables such as the thermal free energy and scalar two-point functions, predictions which could be tested on the lattice.
\end{abstract}
\maketitle

\section{I.\ Introduction}

\noindent
Top-down holography can be extended well beyond AdS/CFT, since the decoupling between the open and closed string degrees of freedom that underpins the correspondence does not require conformal symmetry. For $N$ D$p$-branes with $p\neq3$, holography relates $(p+1)$-dimensional maximally supersymmetric Yang--Mills theory (MSYM) to type II string theory on the near-horizon geometry of the backreacted branes~\cite{Itzhaki:1998dd}. The gauge theory is not conformal, since its 't~Hooft coupling $\lambda=g_{\rm YM}^2N$ has mass dimension $3-p$. This is reflected in the gravitational background which features a running dilaton $\Phi$.  
This running breaks scale invariance only mildly. In the so-called dual frame, $\ds_{\rm dual}^2=\e^{2\Phi/(p-7)}\ds^2_{\rm str}$~\cite{Boonstra:1998mp,Kanitscheider:2008kd}, the metric is ${\rm AdS}_{p+2}\times S^{8-p}$:
\begin{equation}
	\begin{split}
		\ds_{\rm dual}^2&=\ell^2\Big(\tfrac{4}{(5-p)^2}\,\tfrac{\dd z^2+\dd x_{1,p}^2}{z^2}+\dd\Omega_{8-p}^2\Big)\,,\\
		F_{8-p}&=(7-p)\,\ell^{7-p}\vol_{8-p}\,,\\
		\e^{\Phi}&=(z/\ell)^{\eta(7-p)/(2(3-p))}\,,\qquad \eta=\tfrac{(3-p)^2}{5-p}\,,
	\end{split}
	\label{Dpbranes}
\end{equation}
with $\vol_{8-p}$ the volume form on the unit sphere and $(\ell/\ell_s)^{7-p}\propto N$ fixed by flux quantization.\footnote{The running of the dilaton implies that the background \eqref{Dpbranes} as well as other backgrounds discussed in this paper has a limited range of validity controlled by the coordinate $z$ \cite{Itzhaki:1998dd}.} 
This structure means that a constant shift of $\Phi$, i.e. a rescaling of the string coupling constant $g_s$, rescales the supergravity action and thus maps one solution to another. A dilatation transformation $(z,x^\mu)\mapsto t\,(z,x^\mu)$ leaves the dual-frame metric invariant but must be compensated by a dilaton shift to leave the full supergravity background invariant. This is \emph{scale covariance}, the gravitational counterpart of the generalized conformal symmetry of the dual large $N$ gauge theory~\cite{Jevicki:1998yr,Jevicki:1998ub,Kanitscheider:2008kd}, also known as scaling similarity~\cite{Biggs:2023sqw}.\footnote{D5-branes do not lead to a scale-covariant structure of the type we study here and we will not discuss $p=5$ further in this work.} 

This scale-covariant structure is characterized by the scaling exponent $\eta$ which combined with the AdS metric in dual frame can be leveraged to deduce a holographic dictionary for physical observables in the dual QFT. Consider the thermal free energy density computed by evaluating a supergravity action on a thermal black brane background. It can be shown that scale covariance fixes the form of the free energy to be \cite{Itzhaki:1998dd,Peet:1998wn} 
\begin{equation}\label{eq:intro-free-energy}
	\mathcal F \sim \,T^{d+\eta}\,.
\end{equation}
Scalar excitations of the supergravity background~\eqref{Dpbranes} are mapped to local operators in the dual QFT and scale covariance dictates that their 2pt-functions take the form
\begin{equation}\label{eq:2pt}
\langle \mathcal{O}(x)\mathcal{O}(y)\rangle \sim \frac{1}{|x-y|^{2\Delta-\eta}}\,,
\end{equation}
where the scaling dimension $\Delta$ is determined by the mass of the scalar field~\cite{Jevicki:1998yr,Jevicki:1998ub,Kanitscheider:2008kd,Biggs:2023sqw}. Similarly, the 3pt-functions of local operators dual to supergravity modes are also determined in terms of $\eta$ and their scaling dimension~\cite{Biggs:2025qfh,Bobev:2025idz,Bobev:2026xxx}.

Scale covariant holography therefore provides predictions for the strong-coupling behavior of non-conformal planar gauge theories that can be tested. For $p<3$ the gauge theories are super-renormalizable and well suited to lattice simulations. Monte Carlo studies of the D0-brane matrix model confirm the $T^{14/5}$ behavior in~\eqref{eq:intro-free-energy}, as well as the structure of the scalar correlators in~\eqref{eq:2pt}, see~\cite{Anagnostopoulos:2007fw,Catterall:2008yz,Hanada:2008ez,Berkowitz:2016jlq,Hanada:2011fq}. Similarly, lattice studies, see~\cite{Catterall:2017lub,Catterall:2020nmn,Schaich:2025uuo,Joseph:2026fdm}, of 2d and 3d MSYM agree with the dual black brane free energy~\eqref{eq:intro-free-energy}. Supersymmetric localization offers complementary precision tests of scale-covariant holography for supersymmetric observables~\cite{Bobev:2018ugk,Bobev:2019bvq,Bobev:2024gqg}. 

The goal of this work is to demonstrate that the scale-covariant backgrounds in~\eqref{Dpbranes} are only a small corner of a vast landscape of scale-covariant holographic backgrounds in string theory. We provide new explicit examples that arise from branes in string theory with reduced supersymmetry and mirror familiar constructions in AdS/CFT. The scaling exponent $\eta$ is an important characteristic of each of these backgrounds and is in general different from its maximally supersymmetric value in~\eqref{Dpbranes}. We describe three different routes that lead to scale-covariant backgrounds.

\textit{(i) Gauged supergravity.} Just as AdS vacua are critical points of a scalar potential, scale-covariant backgrounds of gauged supergravity solve an algebraic system of equations. We study this system for the 4d electric maximal $\mathrm{ISO}(7)$ gauged supergravity. This theory arises as a consistent $S^6$ truncation of massless IIA supergravity~\cite{Hull:1984yy,Hull:1988jw,Guarino:2015vca}, whose maximally supersymmetric background is the D2-brane background~\eqref{Dpbranes}. A systematic scan uncovers 78 scale-covariant backgrounds with reduced symmetries and different values of $\eta$, including four new supersymmetric ones. All of these backgrounds can be uplifted to solutions of IIA supergravity, which we illustrate explicitly with some examples. In addition, we compute the spectrum of scaling operator dimensions of low-lying supergravity modes and where applicable organize them into supermultiplets. We show that two of the supersymmetric scale-covariant backgrounds are connected by BPS domain walls to the maximally supersymmetric D2-brane solution. These can be interpreted as holographic RG flows dual to mass deformations of 3d MSYM. The algebraic system and all 78 solutions, together with their full mass spectra, supersymmetry properties, BF stability status, and global symmetries, are collected in an ancillary database accompanying this publication and available at \href{https://github.com/Pibom/Scale-covariant-backgrounds}{github.com/Pibom/Scale-covariant-backgrounds}.

\textit{(ii) Wrapped branes.} Placing D-branes on curved manifolds realizes a partial topological twist of the world-volume gauge theory and provides a mechanism to construct supergravity holographic backgrounds~\cite{Maldacena:2000mw,Maldacena:2000yy}. This procedure can be applied to the D-brane backgrounds in~\eqref{Dpbranes} to construct new scale-covariant backgrounds in lower dimensions. We discuss several known wrapped-brane solutions from this perspective~\cite{Edelstein:2001pu,Gomis:2001vk,Boisvert:2024jrl} and construct a new supersymmetric example with dual-frame geometry AdS$_2\times\Sigma_{\mathfrak g}$ dual to a partial topological twist of mass-deformed 3d MSYM.

\textit{(iii) Ten dimensions.} A simple way to construct scale-covariant backgrounds is to replace the $S^{8-p}$ sphere in~\eqref{Dpbranes} by any Einstein manifold of the same curvature. This setup describes the backreaction of a stack of D$p$-branes at the tip of a Ricci-flat cone~\cite{Acharya:1998db}, similar to D3-branes at CY singularities~\cite{Klebanov:1998hh,Morrison:1998cs}. Supersymmetry is preserved when the Einstein space admits Killing spinors, i.e.\ when the cone has special holonomy. For D2-branes these backgrounds can be constructed explicitly by uplifting the gauged supergravity results described above and adapting the analysis of~\cite{Fluder:2015eoa} to find many new $\mathcal N=2$ and $\mathcal N=1$ scale-covariant backgrounds labeled by Sasaki--Einstein (SE) five-manifolds. More generally, we discuss how supersymmetric scale-covariant 10d supergravity backgrounds could be classified using $G$-structure techniques, see~\cite{Tomasiello:2022dwe} for a review.

\section{II. 4d supergravity}
\label{sec:d2-vacua}

\noindent Consider a consistent truncation of 10d type II supergravity to a gauged supergravity in $d+1<10$ dimensions.
The gauged supergravity theory is scale covariant with weight $k$ if it admits a one-parameter rescaling of its fields $\Psi\mapsto\Psi_t$ such that its action transforms as $S[\Psi_t]=t^kS[\Psi]$.  We consider cases where it is possible to employ field redefinitions such that the map acts trivially on all fields except one scalar field; the gauged supergravity dilaton $\varphi$ normalized such that the scaling action is $\varphi\to\varphi+\log t$. This defines the dual frame in the context of the gauged supergravity and is closely related to the dual frame in~\eqref{Dpbranes}, see~\cite{Boonstra:1998mp,Bobev:2026xxx}. Scale covariance dictates that the action takes the form
\begin{equation}
	\begin{aligned}
		S={}& \tfrac{1}{2\kappa^2}\int\!\star \, \e^{k\varphi} \Big[ R-\tfrac12h_1(\partial\varphi)^2-h_{2m}\,\partial_\mu\varphi\,\partial^\mu X^m\\
		&\hspace{7mm}-\tfrac12G_{mn}\,\partial_\mu X^m\partial^\mu X^n
		-\mathcal V(X)+\cdots\Big]\,. 
	\end{aligned}
	\label{eq:prl-covariant-action}
\end{equation}
Here $X$ are the scalar fields on which the scaling transformation acts trivially and $\mathcal V(X)$ is their potential. The ellipsis comprises kinetic terms for fermions, vectors, and form fields as well as Chern--Simons, BF, and other couplings.  For $k\neq0$ the dilaton is a runaway direction, so these theories generically do not admit maximally symmetric solutions with all scalar fields constant. Instead, their Poincar\'e-invariant backgrounds exhibit scale covariance,
\begin{equation}
    	\ds_{d+1}^2 =L^2\Big(\tfrac{\dd z^2 + \dd x_{1,d-1}^2}{z^2}\Big)\,,\qquad \varphi = - \tfrac{\eta}{k}\log \tfrac{z}{L}\,.
  	\label{eq:prl-scaling-ansatz}
\end{equation}
All scalar fields except the dilaton take constant values $X^m=X_*^m$ and all supergravity fields with spin vanish. The equations of motion then reduce to the following algebraic relations, with all functions evaluated at $X_*$,
\begin{equation}
	\begin{aligned}
		\eta ={}& \frac{2k^2}{2k^2+h_1}\,,\\
		\frac1{L^2} ={}& - \frac{\mathcal V} {(d+\eta)(d+\eta-1)}\,,\\
		0={}& L^2\partial_m\mathcal V + \frac{\eta^2}{2k^2}\partial_mh_1 - \frac{\eta}{k}(d+\eta)h_{2m}\,.
	\end{aligned}
	\label{eq:prl-scaling-equations}
\end{equation}
The first two relations follow from the Einstein and dilaton equations, the third follows from the remaining scalar equations and determines their fixed values $X_*$.  Notice that scale-covariant backgrounds are not generally critical points of the scalar potential $\mathcal V$. 

Once a background has been found, we can linearize the equations of motion around it. After a subtle unmixing~\cite{Kanitscheider:2008kd,Kanitscheider:2009as,Bobev:2026xxx}, we find that the linearized modes behave as massive fields propagating in AdS of effective spacetime dimension $d+1+\eta$. Using this we can assign a notion of mass to each mode and use the usual holographic dictionary, i.e. $m^2L^2 = \Delta(\Delta-d-\eta)$ for scalars, to translate this to a scaling dimension of the dual QFT operator and thus deduce its 2pt-function as in~\eqref{eq:2pt}. Moreover, we can use the BF criterion, $m^2L^2\geq-(d+\eta)^2/4$, to deduce the perturbative stability of a given background.

\paragraph{4d $\mathrm{ISO}(7)$ supergravity.}
We now apply the general procedure described above to find scale-covariant solutions in a consistent truncation of type IIA supergravity on $S^6$. The 10d metric and form fields reduce to a 4d metric, 70 scalar fields, and 28 vector fields. Together with fermions, these arrange into a 4d maximal supergravity in which the vector fields gauge the non-abelian $\mathrm{ISO}(7)$ algebra~\cite{Hull:1984yy,Hull:1988jw}. We remark that since the Romans mass is zero, the 4d theory is \emph{electrically} gauged and exhibits scale covariance. 

Due to the unwieldy number of scalar fields, we follow the approach taken in \cite{Bobev:2020qev} and truncate the 4d theory to its $\mathbb Z_2^3$ and $\mathbb Z_2^2$-invariant subsectors. This reduces the number of scalar fields from 70 to 14 and 22, respectively. The 14-scalar model can be described in ${\cal N}=1$ supergravity language in terms of a K{\"a}hler potential and a holomorphic superpotential\footnote{The Einstein frame Lagrangian is $\mathcal L = R - 2K_{i\bar\jmath}\partial z^i\partial\bar z^{\bar\jmath} - V$, where $V=\e^K (K^{i\bar{j}}D_i \mathcal{W}D_{\bar j}\overline{\mathcal{W}} - 3|\mathcal{W}|^2 ) = \e^{2\varphi}\mathcal{V}$.}
\begin{equation}\label{Kaehlersuper}
\begin{split}
K &= -\sum_{i=1}^7\log[2\Im z_i]\,,\\
{\cal W} &= 4 g \big(z_1 z_2 z_3+z_4 z_5 z_3+z_6 z_7 z_3+z_2 z_4 z_6\\
&\qquad\qquad\qquad\qquad+z_1 z_5 z_6+z_1 z_4 z_7+z_2 z_5 z_7\big)\,,
\end{split}
\end{equation}
where $g$ is the 4d gauge coupling constant. To separate the 4d dilaton field and transform the Lagrangian in dual frame we perform the field redefinition $z_i = \e^{-2\varphi} \hat z_i$ and the Weyl rescaling $g_{\mu\nu} = \e^{2\varphi}g_{\mu\nu}^\text{Einst}$. The 22-scalar model exhibits the same structure but is considerably more complicated to write down. We include an ancillary file with this submission that contains the relevant parts of the bosonic action~\eqref{eq:prl-covariant-action} with $k=-2$.

Equipped with the explicit action we can use the equations in~\eqref{eq:prl-scaling-equations} to set up a numerical search for solutions. We have identified a total of 78 scale-covariant backgrounds of the 22-scalar model. We label the solutions by the symbol \texttt{Z} followed by the first six digits after the decimal point of the numerical value of $Lg$. In Table~\ref{tab:prl-stable-vacua} we present the six BF stable solutions, five of which are supersymmetric, together with the corresponding values of the scaling exponent $\eta$. This pattern mirrors the dyonically gauged $\mathrm{ISO}(7)_c$ AdS backgrounds, whose large non-supersymmetric landscape is dominated by BF unstable extrema~\cite{Bobev:2020qev}. As it turns out 72 of the solutions, including all in the table, can be found in the 14-scalar model. The maximally supersymmetric background, denoted as \texttt{Z666666} in the table, uplifts to the type IIA background in~\eqref{Dpbranes} for $p=2$. The uplifts of all other solutions are deformations of the background in~\eqref{Dpbranes} which involve a squashing of the internal $S^6$, a warp factor in the metric, and additional non-trivial form fields. In the Supplemental Material we present explicit examples of such uplifted solutions. In addition, for the five supersymmetric backgrounds, we present a detailed analysis of the spectrum of excitations of the four-dimensional supergravity fields and their organization into supermultiplets.

\begin{table}[t]
	\caption{BF stable scale-covariant backgrounds of the electric 4d $\mathrm{ISO}(7)$ supergravity.  The values of $\eta$ for the last two points are roots of a degree-14
		polynomial.}
	\label{tab:prl-stable-vacua}
	\centering
	\small
	\setlength{\tabcolsep}{4pt}
	\begin{ruledtabular}
		\begin{tabular}{crcc}
			Label    & $\eta$                             & Symmetry         & SUSY           \\
			\hline
			\texttt{Z666666}    & $\tfrac13$                         & $\mathrm{SO}(7)$ & $\mathcal N=8$ \\
			\texttt{Z597968}    & $\tfrac1{24}(\sqrt{1569}-33)$      & $\mathrm{SU}(3)$ & $\mathcal N=1$ \\
			\texttt{Z596827}    & $\tfrac14(\sqrt{65}-7)$            & $\mathrm G_2$            & $\mathcal N=0$ \\
			\texttt{Z570472}    & $\tfrac16(\sqrt{73}-7)$            & $\mathrm{U}(2)$  & $\mathcal N=2$ \\
			\texttt{Z448357} & $0.1649599992$                     & $\mathrm{U}(1)$  & $\mathcal N=1$ \\
			\texttt{Z394258} & $0.1433927955$                     & $\mathrm{U}(1)$  & $\mathcal N=1$
		\end{tabular}
	\end{ruledtabular}
\end{table}

\begin{figure}[t]
	\centering
	\includegraphics[width=\columnwidth]{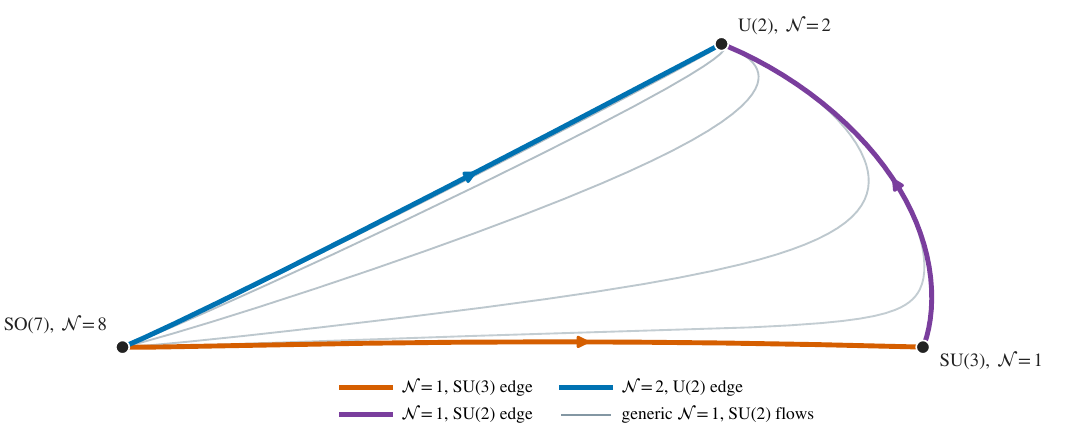}
	\caption{Supersymmetric RG flows between \texttt{Z666666}, \texttt{Z597968}, and \texttt{Z570472}. Thin curves
		sample the $\mathcal N=1$, $\mathrm{SU}(2)$ family of flows; arrows point from UV to IR.}
	\label{fig:prl-triangle-flows}
\end{figure}

\paragraph{Supersymmetric flows.}
We now focus on three scale-covariant backgrounds and the BPS domain wall flows between them: the $\mathcal N=8$, $\mathrm{SO}(7)$ point \texttt{Z666666}; the $\mathcal N=1$, $\mathrm{SU}(3)$ point \texttt{Z597968}; and the $\mathcal N=2$, $\mathrm{U}(2)$ point \texttt{Z570472}.  The direct $\mathcal N=8\to \mathcal N=2$ flow preserves $\mathcal N=2$ supersymmetry and $\mathrm{U}(2)$ bosonic symmetry. A unique flow preserves $\mathrm{SU}(3)$ and flows between the $\mathcal N=8$ and the $\mathrm{SU}(3)$ $\mathcal N=1$ points. All other flows are $\mathrm{SU}(2)$ invariant and preserve $\mathcal N=1$ supersymmetry which is enhanced at the end of the flow to $\mathcal N=2$. Note that although we use familiar RG-flow language to describe these solutions, all \emph{fixed points} represent scale-covariant backgrounds that break conformal invariance. The flow solutions are most conveniently described in terms of an $\mathrm{SU}(2)$ invariant truncation of the 14 scalar model which is obtained by setting $z_3=z_1$, $z_6=z_4$, and $z_7=z_5$ in~\eqref{Kaehlersuper}. The $\SU(3)$ invariant flow can be treated separately by further setting $z_5=z_4$ and $z_2=z_1$. The domain wall solutions are solutions to the BPS equations of the 4d supergravity which reduce to a system of ODEs controlled by the real superpotential $W = \e^{K/2} |\mathcal{W}|$  
\begin{equation}
K_{\bar\jmath i}( z^{i})' \!=\! -  \sqrt 2\partial_{\bar\jmath}W\,,~~ K_{i\bar\jmath}(\bar z^{\bar\jmath})' \!=\!-  \sqrt 2\partial_{i}W\,,~~ A' \!=\! \frac{W}{\sqrt 2}\,.
\end{equation}
Here primes denote $r$-derivatives, the K{\"a}hler metric is $K_{i\bar\jmath } = \partial_i \partial_{\bar\jmath}K$ and the Einstein frame metric is written as $\ds_4^2 = \dd r^2 + \e^{2A}\dd x_{d}^2$. After performing the field redefinition discussed below~\eqref{Kaehlersuper} to isolate the dynamics of the dilaton, it is straightforward to numerically construct flows between the three scale-covariant backgrounds. We have constructed a number of numerical solutions by shooting from the $\mathcal N=2$, $\mathrm{U}(2)$ point \texttt{Z570472} point `up' towards the maximally supersymmetric background. The results are sketched in Fig.~\ref{fig:prl-triangle-flows} and will be discussed further in upcoming work~\cite{Bobev:2026nch}.

\paragraph{Field-theory RG flows.}
The supergravity RG flows discussed above can be interpreted using the Lagrangian description of the dual 3d MSYM theory. This weakly coupled description is not expected to capture the details of the strongly coupled dynamics probed by the supergravity solutions but it is nevertheless informative to organize the possible deformations of 3d MSYM. It is convenient to work in 3d $\mathcal N=1$ superspace where the theory is described by a vector multiplet and 7 real adjoint scalar superfields $X^I$. We can organize six of these superfields into the complex linear combinations $Z^a=X^{2a-1}+\mathrm i X^{2a}$, $a=1,2,3$.  The two-dimensional space of supersymmetric relevant deformations spanned by the holographic RG flows above is described by the $\mathcal N=1$ superpotential
\begin{equation}\label{eq:prl-two-mass-deformation}
	\mathscr W={}  \mathscr W_0 +M_1\!\left[Z^a\bar Z^{\bar a}-6(X^7)^2\right] +\operatorname{Re}\!\left[M_2(Z^3)^2\right]\,,	
\end{equation}
where $\mathscr W_0$ is the maximally supersymmetric cubic superpotential~\cite{Mauri:2008ai,Buchbinder:2010ez} and $M_{1,2}$ are real mass parameters. For $M_2=0$ the flow preserves $\mathrm{SU}(3)$ global symmetry and the dual gravity flow is the orange line in Fig.~\ref{fig:prl-triangle-flows}, while for $M_1=0$ supersymmetry is enhanced to $\mathcal{N}=2$, the global symmetry is $\mathrm{U}(2)$, and the holographic dual is the blue line in Fig.~\ref{fig:prl-triangle-flows}. For generic values of $M_{1,2}$ the global symmetry is $\mathrm{SU}(2)$ and the holographic dual is the family of $\mathcal{N}=1$ supergravity flows in Fig.~\ref{fig:prl-triangle-flows}. A very similar family of 3d $\mathcal{N}=1$ holographic RG flows exists for the 3d ABJM SCFT on the worldvolume of M2-branes~\cite{Bobev:2009ms}.

\section{III. Wrapped branes}
\label{sec:wrapped-branes}

\noindent Another route to scale-covariant backgrounds is to wrap D$p$-branes on compact cycles and study their backreaction. In general there are no (conformal) Killing spinors on curved spaces but one can preserve some supersymmetry by employing a partial topological twist, i.e. turning on an R-symmetry background field to cancel the spin connection~\cite{Maldacena:2000mw,Maldacena:2000yy}.  Applied to the D$p$-brane backgrounds in~\eqref{Dpbranes}, this yields gravity duals of partially twisted Yang--Mills theories. As for wrapped D3- and M5-branes, different R-symmetry embeddings and flavor fluxes give inequivalent supergravity backgrounds~\cite{Benini:2013cda,Bah:2012dg}, while the universal twist uses only the R-symmetry of the preserved supersymmetry~\cite{Maldacena:2000mw,Benini:2015bwz,Bobev:2017uzs}.

\textit{Riemann-surface twists.}
The D$p$-brane solutions on
$\Sigma_{\mathfrak g}$ for $p=2,4,6$ were constructed in~\cite{Boisvert:2024jrl}. They can be written in a uniform way as solutions of the STU model of an appropriate $(p+2)$-dimensional gauged supergravity. In dual frame the metric and gauge fields read
\begin{align}
	\ds_{p+2}^2= L_p^2 \tfrac{\dd z^2+\dd x_{1,p-2}^2}{z^2} + R_p^2\ds_{\Sigma_{\mathfrak g}}^2\,, \quad
	F^{I}= \mathfrak p_I\operatorname{vol}_{\Sigma_{\mathfrak g}}\,, 
	\label{eq:uniform-wrapped-branes}
\end{align}
with respectively $3,2,1$ Cartan gauge fields and scalar profiles
\begin{equation}
	\begin{array}{c@{:\quad}l}
		\mathrm{D2} & \e^{\varphi}=c_0z^{\tfrac16},\quad
		\e^{x_I/2}=c_I,\quad I=1,2,3,      \\
		\mathrm{D4} & \e^{\varphi}=c_0z^{-1},\quad
		\e^{2\lambda_I}=c_I,\quad I=1,2,            \\
		\mathrm{D6} & \e^{\varphi}=c_0z^{\tfrac32},\quad
		\e^{6\lambda}=c_1 .
	\end{array}
	\label{eq:uniform-wrapped-scalars}
\end{equation}
The constants and the magnetic charges obey algebraic relations, including the topological twist constraint, which can be found in~\cite{Boisvert:2024jrl}. The D2 and D4 backgrounds exist for all three curvature signs of $\Sigma_{\mathfrak g}$, while the D6 solutions are regular only for $\Sigma_{\mathfrak g}=S^2$.  The spindle and disk solutions of~\cite{Boisvert:2024jrl} yield scale-covariant backgrounds in an analogous way. We note in passing that the D4-brane solutions in~\eqref{eq:uniform-wrapped-branes}-\eqref{eq:uniform-wrapped-scalars} can also be obtained as a circle reduction of the AdS$_5$ backgrounds of 11d supergravity constructed in~\cite{Bah:2011vv,Bah:2012dg} which arise from twisted compactifications of M5-branes on $\Sigma_{\mathfrak{g}}$.

The wrapped brane solutions can be uplifted to 10d supergravity and by analyzing the running dilaton one finds that the parameter $\eta$ is the same as for the parent flat brane solutions in~\eqref{Dpbranes}, namely $\eta=(p-3)^2/(5-p)$. This is perhaps not surprising; in the 4d supergravity D2-brane solutions presented above, axions played a crucial role in changing the value of $\eta$. This is because they appear in the dilaton kinetic function $h_1$ in~\eqref{eq:prl-covariant-action}. Since the wrapped brane solutions do not feature axions, and the kinetic term for the gauged supergravity dilaton is canonical, we should expect that $\eta$ is unmodified.

\textit{A supersymmetric wrapped} \texttt{Z570472} \textit{background.}
A partial topological twist can also be applied to the $\mathcal N=2$ background \texttt{Z570472} by turning on magnetic flux for the $\mathrm U(1)$ R-symmetry subgroup of the $\mathrm U(2)$ global symmetry. We have constructed a numerical flow in four-dimensional electric $\mathrm{ISO}(7)$ gauged supergravity preserving two real supercharges. It connects the unwrapped scale-covariant background to an infrared scale-covariant geometry with dual-frame metric $\mathrm{AdS}_2\times\Sigma_{\mathfrak g}$. The axionic scalars run along the flow, yielding an infrared exponent $\eta=0.270048\ldots$ that differs from the parent value in Table~\ref{tab:prl-stable-vacua}.

\textit{Non-abelian twists.}
Cycles of dimension greater than two require embedding a non-abelian spin connection, or the reduced holonomy group, into the R-symmetry.  The holographic dual to MSYM in seven dimensions is governed by eight-dimensional $\mathrm{SU}(2)$ gauged supergravity, and identifying that $\mathrm{SU}(2)$ with the spin group of an $S^3$ preserves four supercharges. The IR dual QFT is 4d pure $\mathcal N=1$ SYM, together with the Kaluza--Klein tower on the $S^3$~\cite{Edelstein:2001pu,Gomis:2001vk,Atiyah:2000zz}.  The corresponding scale-covariant supergravity solution can be written in dual frame as
\begin{equation}
	\begin{aligned}
		\ds_8^2    & =L^2\tfrac{\dd z^2+\dd x_{1,3}^2}{z^2}+\frac{L^2}{3}\dd\Omega_3^2, \\[-2pt]
		\e^{2\varphi} & =\frac{g^3z^3}{36},\qquad
		F^a=-\frac{1}{8g}\epsilon^{abc}w_b\wedge w_c ,
	\end{aligned}
	\label{eq:prl-d6-three-cycle}
\end{equation}
with $\dd\Omega_3^2=\frac14\sum_a w_a^2$ written in terms of left-invariant one-forms of $\mathrm{SU}(2)$.  As before, the scaling exponent $\eta$ is the same as for the parent flat D6-brane theory. There are other possible non-abelian twists associated with D6-branes wrapping 4-manifolds. Wrapping the branes on a co-associative 4-cycle in a G$_2$ manifold leads to a 3d $\mathcal{N}=1$ SYM theory at low energies and the dual 8d supergravity solution in dual frame has an $\mathrm{AdS}_4\times S^4$ metric~\cite{Hernandez:2001bh}. A related $\mathrm U(1)$ twist on K\"ahler four-cycles in a Calabi--Yau threefold yields three-dimensional $\mathcal N=2$ SYM instead and the corresponding scale-covariant supergravity background was presented in~\cite{Gomis:2001vg}.

For D4-branes the R-symmetry is $\mathrm{SO}(5)$ which is large enough to accommodate the $\mathrm{SO}(3)$ and $\mathrm{SO}(4)$ spin connections of three- and four-cycles. The corresponding supergravity solutions can be constructed by a circle reduction of the known AdS$_4$ and AdS$_3$ vacua of 11d supergravity obtained by a partial topological twist of M5-branes on three- or four-cycles, respectively, see~\cite{Bobev:2017uzs} for a summary of these backgrounds. 

\section{IV. 10d supergravity}
\label{sec:ten-dimensional-vacua}

\noindent Consistent truncations to gauged supergravity are expected to capture only a small corner of the 10d scale-covariant landscape. A simple extension of~\eqref{Dpbranes} comes from backreacting D$p$-branes at the tips of Ricci-flat cones $C(Y_{8-p})$, yielding near-horizon dual-frame geometries of the form $\mathrm{AdS}_{p+2}\times Y_{8-p}$ with $R_{ij}=(7-p)g_{ij}$. The dilaton, and hence $\eta$, remain unchanged. Supersymmetry may be preserved when $Y_{8-p}$ admits real Killing spinors, equivalently when the cone has special holonomy~\cite{Acharya:1998db}. For example, D1-branes on CY$_4$ or $\mathrm{Spin}(7)$ cones yield scale-covariant backgrounds with 2d $\mathcal{N}=(0,2)$ or $\mathcal{N}=(0,1)$ supersymmetry, respectively. Explicit examples use the 7d Sasaki--Einstein manifolds of~\cite{Gauntlett:2004hh} or the squashed $S^7$ with Einstein metric~\cite{Awada:1982pk}. 

The same idea can be used to deform the backgrounds of Sec.~II. The $\mathrm{SU}(3)$-invariant ones can be uplifted to 10d massless type IIA supergravity~\cite{Guarino:2015vca,Varela:2015uca}. The resulting solutions are determined by the four-dimensional scalars together with the Sasaki--Einstein data of $S^5$, as detailed in the Supplemental Material. We can simply replace the SE data of $S^5$ with those of any SE manifold to generate infinite families of new backgrounds~\cite{Fluder:2015eoa}. For the $\mathcal N=8$ background this yields a family that generically preserves $\mathcal N=2$, with enhancement to $\mathcal N=8$ only at $Y_5=S^5$. For the supersymmetric $\mathrm{SU}(3)$-invariant background \texttt{Z597968}, the additional scalars deform the metric and turn on further fluxes, so that this family preserves only $\mathcal N=1$ supersymmetry. 

More generally, one may apply $G$-structure techniques, see~\cite{Tomasiello:2022dwe} for a review, to systematically classify supersymmetric scale-covariant backgrounds of 10d supergravity.  We briefly sketch how this classification may proceed for backgrounds of type IIA supergravity relevant for D2-branes. Consider the most general ansatz for a scale-covariant background,
\begin{equation}
	\begin{aligned}
		\ds^2_{\rm dual}&=\e^{2A}L^2\,\tfrac{\dd z^2+\dd x^2_{1,2}}{z^2} + \ds^2(M_6)\,,\\
		\Phi&=\phi+\tfrac{5\eta}{2}\log z\,,
	\end{aligned}
\end{equation}
with $A$ and $\phi$ functions on $M_6$. Scale covariance then dictates the radial dependence of the fluxes: under $(z,x^\mu)\mapsto t\,(z,x^\mu)$, the NS-NS and R-R fluxes scale as
\begin{equation}\label{eq:HFtscaling}
	H_3\mapsto t^{\eta}H_3\,,\qquad
	F_n\mapsto t^{(n-6)\eta/2}F_n\,.
\end{equation}
Therefore a consistent ansatz for the fluxes is to take an appropriate differential form on $M_6$ multiplied by a fixed power of $z$. Note that the Page six-form, $\widehat{F}_6 = F_6-B_2\wedge F_4 +\frac12 B_2\wedge B_2\wedge F_2$, used for D2-brane flux quantization has weight zero which ensures that the number of D2-branes $N$ is not affected by the scaling transformation. Since the dilaton runs, the string frame background is a warped $\mathbb R^{1,2}\times M_7$ space with $M_7=\mathbb R_z\times M_6$. The supersymmetry conditions for these backgrounds should therefore be the same as for a Minkowski$_3$ compactification with an additional constraint that ensures the homogeneity in $z$. As usual the Killing spinor equations can be traded for first-order identities for spinor bilinears. Stripping off the $z$-dependence with the weights in~\eqref{eq:HFtscaling} turns these into torsion conditions on $M_6$ which, together with the Bianchi identities and the flux equations of motion, should provide a system of differential equations that fixes the possible backgrounds. Setting $\eta=0$ should reduce this system to the one arising from a classification of supersymmetric AdS$_4$ solutions of type IIA supergravity, see for example~\cite{Lust:2004ig}.

\section{V. Discussion}
\label{sec:discussion}
%

\noindent Our analysis opens up many questions. Consistent truncations to gauged supergravities associated to D$p$-branes with $p\neq2$ can be utilized to systematically scan for scale-covariant backgrounds as we did for the 4d $\mathrm{ISO}(7)$ theory. ExFT methods should be developed to systematically compute the KK spectra of these supergravity solutions, see~\cite{Samtleben:2025fta} for a review, and the $G$-structure conditions of Sec.~IV should be solved to classify scale-covariant backgrounds directly in ten dimensions. On the QFT side, supersymmetric localization and lattice simulations should be pushed further to test our predictions for the thermal free energy and for two-point functions of local operators. The many perturbatively unstable non-supersymmetric backgrounds we find raise the question whether a Swampland criterion like~\cite{Ooguri:2016pdq} forbids them. The sole exception, the $G_2$-invariant background of Table~\ref{tab:prl-stable-vacua}, therefore deserves closer scrutiny: its massive IIA analogue is stable at the full KK level~\cite{Guarino:2020flh} yet decays through a bubble of nothing~\cite{Bomans:2021ara}. Finally, it remains to understand the dynamical reason for the emergence of scale covariance in holographic strongly coupled planar gauge theories and whether it survives beyond the supergravity approximation. 

\smallskip

\noindent \textbf{Acknowledgements:}~We would like to thank Thomas Fischbacher, Guillermo Mera \'Alvarez, Siyul Lee, Juan Maldacena, Joe Minahan, Anton Nedelin, Hynek Paul, Krzysztof Pilch, Anayeli Ram\'irez, and Kostas Skenderis for useful discussions. NPB is supported by the FWO projects G003523N, G094523N, and G0E2723N, and the KU Leuven C1 project C16/25/01. PB acknowledges support from the ERC Consolidator Grant No. 101044226 ``Exact Results from Broken Symmetries'' (BrokenSymmetries). FFG is supported in part by the Science and Technology Facilities Council (Consolidated Grant ST/C004240/1).

AI was used to verify some of the algorithms used to obtain the results presented in this paper. The authors take full responsibility for the content of the paper.

\smallskip

\noindent\textit{Note added:} While we were completing this manuscript we became aware of related work in~\cite{Oviedo:2026} that may overlap with some of our results. We have coordinated the submissions of our manuscripts with the authors of~\cite{Oviedo:2026}.

\bibliography{ScalingVacua}
\bibliographystyle{utphys}

\clearpage
\onecolumngrid
\setcounter{equation}{0}
\renewcommand{\theequation}{S\arabic{equation}}
\renewcommand{\theHequation}{S.\arabic{equation}}
\setcounter{section}{0}
\renewcommand{\thesection}{S\arabic{section}}
\renewcommand{\theHsection}{S.\arabic{section}}

\section*{Supplemental Material}

\section{Type-IIA uplifts of the \texorpdfstring{$\mathrm{SU}(3)$}{SU(3)}-invariant scale-covariant backgrounds}
\label{sec:supp-z0-z4-uplifts}

\noindent In the $\mathrm{SU}(3)$-invariant truncation of 4d $\mathrm{ISO}(7)$ gauged supergravity~\cite{Guarino:2015qaa}, we can solve the equations for scale-covariant backgrounds analytically and prove that the six solutions provide a complete classification of scale-covariant backgrounds in this sector. Each preserves at least an $\mathrm{SU}(3)$ subgroup of the gauge group. Using the uplift formulae of~\cite{Guarino:2015vca,Varela:2015uca}, we construct the corresponding type IIA supergravity backgrounds, which we present explicitly below.

The metric of the 6d internal manifold can be viewed as a deformation of the sine cone over $S^5$. This in turn allows for a generalization of these solutions to the sine-cone over any 5d Sasaki--Einstein manifold $Y_5$ along the lines of~\cite{Fluder:2015eoa}. The local structure of $Y_5$ can be compactly described in terms of the metric and a set of canonical differential forms
\begin{gather}
	\ds^2(Y_5)=\ds_{\rm KE}^2+\theta^2,\qquad
	\dd\theta=2J,\qquad \dd J=0,\nonumber\\
	\dd\Omega=3\mathrm i\,\Omega\wedge\theta,\qquad
	J\wedge\Omega=0,\qquad
	\Omega\wedge\bar\Omega=2J\wedge J .
	\label{eq:supp-se5-structure}
\end{gather}
Here $\ds_{\rm KE}^2$ is the metric on a K\"ahler-Einstein manifold with K\"ahler form $J$ and $\theta$ is the 1-form dual to the Reeb vector. For the choice $Y_5=S^5$ we find that $\ds_{\rm KE}^2$ is the Fubini-Study metric on $\mathbb{CP}^2$ and $\theta$ is the Hopf fiber 1-form.

As described in~\cite{Guarino:2015qaa} the $\rm SU(3)$-invariant sector of the 4d gauged supergravity comprises six real scalar fields. For the scale-covariant backgrounds of interest here two of these fields can be set to zero by a gauge choice. The 4d solutions then take the familiar form in Einstein frame
\begin{equation}
	\e^{\varphi_4} =\e^{-\frac{\eta\,r}{2L}}\,,      \qquad
	\ds^2_{4,\rm E}  =\e^{-2\varphi_4}\left[\dd r^2
	+\e^{2r/L}
	\dd x_{1,2}^2\right]\,,
	\label{eq:supp-uplift-running}
\end{equation}
where $\varphi_4$ is the 4d dilaton and the other five scalar fields are given by
\begin{equation}
	\e^{\phi_4}=\e^{2\varphi_4}\sqrt{\bar x},\qquad
	\chi_4=\e^{-2\varphi_4}\bar\chi,\qquad
	\widetilde\zeta_4=2\e^{-2\varphi_4}\bar\rho,\qquad
	\zeta_4=a_4=0.
\end{equation}
We mostly follow the notation for the scalar fields in Section 3.1 of~\cite{Guarino:2015qaa} but add a subscript $4$ to all scalars.\footnote{Our dilaton differs by a factor of two from the one used in \cite{Guarino:2015qaa}, $2\varphi_4=\varphi_{\rm GV}$.} Here $L$ and $\eta$ are the AdS radius and the scaling exponent defined in~\eqref{eq:prl-scaling-equations} and we have $k=-2$ as appropriate for D2-branes.

To specify the 10d background we first define
\begin{gather}
	\mathcal X = 1 + \bar\chi^2\,,\qquad
	\mathcal Y = 1 + \bar x\bar\rho^2\,,\nonumber\\
	D_1 = \mathcal Y\sin^2\xi + \bar x\mathcal X\cos^2\xi,\qquad
	D_2 = \sin^2\xi + \bar x\cos^2\xi\,,\nonumber\\
	\mathcal P = \e^{\frac{3}{2}\varphi_4}\bar x^{1/8}\mathcal X^{1/4}
	D_1^{1/2} D_2^{1/8}\,.
	\label{eq:supp-uplift-functions}
\end{gather}
The 10d Einstein-frame metric and dilaton are given by
\begin{align}
	\ds^2_{10,\rm E} = {} & \mathcal P\bigg\{\ds^2_{4,\rm E}
	+\frac1{g^2\e^{2\varphi_4}}\bigg[
	\frac{\dd\xi^2}{\bar x\mathcal X}
	+\sin^2\xi\left(\frac{\ds_{\rm KE}^2}{D_1}
	+\frac{\theta^2}{\mathcal X D_2}\right)\bigg]\bigg\}\,,
	\label{eq:supp-uplift-metric}                                        \\
	\e^{\Phi}={} & \e^{5\varphi_4}\bar x^{3/4}\mathcal X^{-1/2}
	D_1^{-1}D_2^{3/4}\,.
	\label{eq:supp-uplift-dilaton}
\end{align}
Writing $\Omega=\Omega_R+\mathrm i\Omega_I$ we find that the NS-NS and R-R potentials can be written as
\begin{align}
	B_2={} & \frac{\e^{2\varphi_4}}{g^2}\bigg[
	\frac{\bar\chi}{\mathcal X}\sin\xi\,\dd\xi\wedge\theta
	+\frac{\bar x\bar\chi}{D_1}\sin^2\xi\cos\xi\,J
	+\frac{\bar x\bar\rho}{D_1}\sin^3\xi\,\Omega_R\bigg]\,,
	\label{eq:supp-uplift-b2}\\
	C_1={} & \frac{\bar\chi^2-\bar x\bar\rho^2}{g\e^{4\varphi_4}D_2}
	\sin^2\xi\cos\xi\,\theta\,,
	\label{eq:supp-uplift-c1}\\
	C_3={} & \cos^2\xi\,\mathcal C^0_3+\sin^2\xi\,\mathcal C^1_3
	-\frac{\bar\chi\mathcal Y}{g^3\e^{2\varphi_4} D_1}
	\sin^4\xi\,J\wedge\theta\nonumber\\
	& +\frac{\bar\rho}{g^3\e^{2\varphi_4}}\sin^2\xi\,
	\Omega_I\wedge\dd\xi
	+\frac{\bar x\mathcal X\bar\rho}{g^3\e^{2\varphi_4} D_1}
	\sin^3\xi\cos\xi\,\Omega_R\wedge\theta \,.
	\label{eq:supp-uplift-c3}
\end{align}
The components of the 3-form along the non-compact space-time directions are given by
\begin{align}
	\mathcal C^I_3={} & -\frac{gL\,Q_I}{3+\eta}
	\e^{(3+\eta)\,r/L}\omega_3\,,
	\label{eq:supp-external-c3}                                                                      \\
	Q_0={} & \frac{\mathcal X}{2}\left[
	12\bar x-2\bar x^2\mathcal X^2-12\bar x^2\bar\chi^2\bar\rho^2\right]\,,
	\nonumber\\
	Q_1={} & \frac12\left[8+2\bar x\mathcal X
	+4\bar x\bar\rho^2(1-3\bar\chi^2)
	-2\bar x^2\bar\rho^2\bar\chi^2\mathcal X
	-4\bar x^2\bar\rho^4(1+3\bar\chi^2)\right]\,,
	\label{eq:supp-uplift-q}
\end{align}
where $\omega_3=\dd x^0\wedge\dd x^1\wedge\dd x^2$. The type IIA supergravity field strengths are
\begin{equation}
	F_0=0,\qquad H_3=\dd B_2\,,\qquad F_2=\dd C_1\,,\qquad
	F_4=\dd C_3+C_1\wedge H_3\,.
	\label{eq:supp-uplift-strengths}
\end{equation}

With the definitions above the six $\rm SU(3)$-invariant scaling solutions are fully determined in terms of the constants $(\bar x,\bar{\chi},\bar{\rho})$ which in turn fix $L$ and $\eta$. We present the explicit values of these constants in Table~\ref{tab:supp-su3-uplifts}.

\begin{table}[htb]
	\caption{The six $\mathrm{SU}(3)$-invariant scale-covariant backgrounds.  The symmetry column gives the full residual symmetry for $Y_5=S^5$. Here $b_\star\approx 0.153363$ is the real root of $b_\star^3-9b_\star^2+47b_\star-7=0$.}
	\label{tab:supp-su3-uplifts}
	\centering
	\renewcommand{\arraystretch}{1.6}
	\begin{ruledtabular}
		\begin{tabular}{cccccc}
			& Symmetry & $\bar x$ & $\bar\chi$ & $\bar\rho$  & $\eta$ \\
			\hline
			\texttt{Z666666} & $\mathrm{SO}(7)$ & $1$ & $0$ & $0$ & $\frac{1}{3}$\\[3pt]
			\texttt{Z642476} & $\mathrm{SO}(6)$ & $4$ & $0$ & $0$ & $\frac{1}{3}$ \\[3pt]
			\texttt{Z597968} & $\mathrm{SU}(3)$ & $\dfrac{111+3\sqrt{1569}}{100}$ & $-\sqrt{\dfrac{-54+2\sqrt{1569}}{105}}$ & $\sqrt{\dfrac{321-8\sqrt{1569}}{70}}$ & $\dfrac{\sqrt{1569}-33}{24}$\\[7pt]
			\texttt{Z596827} & $G_2$ & $1$ & $\sqrt{\dfrac{\sqrt{65}-5}{14}}$ & $\sqrt{\dfrac{\sqrt{65}-5}{14}}$ & $\dfrac{\sqrt{65}-7}{4}$\\[7pt]
			\texttt{Z524611} & $\mathrm{SU}(3)$ & $\dfrac{19-\sqrt{97}}{4}$ & $0$ & $\sqrt{\dfrac{5+2\sqrt{97}}{66}}$ & $\dfrac{\sqrt{97}-9}{4}$\\[7pt]
			\texttt{Z520716} & $\mathrm{SU}(3)$ & $\dfrac{5-b_\star}{2}$ & $0$ & $\sqrt{\dfrac{3-b_\star}{8}}$ & $\dfrac{1-b_\star}{4}$\\[3pt]
		\end{tabular}
	\end{ruledtabular}
\end{table}

Two of the six solutions in Table~\ref{tab:supp-su3-uplifts} preserve supersymmetry. For \texttt{Z666666} we have $Q_0=Q_1=5$ and $gL = 2/3$ and the 10d background simplifies to
\begin{align}
	\ds^2_{10,\rm E}={} &\e^{-\frac{1}{2}\varphi_4}\bigg[
	\dd r^2+\e^{3gr}
	\ds^2(\mathbb R^{1,2})
	+\frac1{g^2}\bigl(\dd\xi^2
	+\sin^2\xi\,\ds^2(Y_5)\bigr)\bigg],\nonumber \\
	\e^{\varphi_4}={}                 & \e^{-gr/4},
	\qquad \e^{\Phi}=\e^{5\varphi_4}, \qquad
	B_2= C_1=0,\qquad
	C_3=-\e^{5gr}\omega_3\,.
	\label{eq:supp-z0-uplift}
\end{align}
This is simply the $\mathcal{N}=8$ D2-brane solution. If we replace $S^5$ in the sine cone with another Sasaki--Einstein manifold supersymmetry is broken to $\mathcal{N}=2$. The background \texttt{Z597968} preserves $\mathcal{N}=1$ supersymmetry for any choice of Sasaki--Einstein data. All background fields in Eqs.~\eqref{eq:supp-uplift-b2}--\eqref{eq:supp-uplift-c3} are turned on for this background and the internal metric is a warped and fiber-squashed sine cone. For $Y_5=S^5$ with its round metric, the internal geometry of all six solutions in Table~\ref{tab:supp-su3-uplifts} is smooth. For a general Sasaki--Einstein manifold $Y_5$, the endpoints are locally cones over $Y_5$ and generically have conical singularities. Note that the warping and fiber-squashing do not introduce new singularities compared to the undeformed Sasaki--Einstein manifold.

\section{Multiplet structure of supersymmetric scale-covariant backgrounds}
\label{sec:supp-susy-multiplets}

\noindent We have explicitly computed the mass spectrum of the 4d supergravity modes for each of the scale-covariant backgrounds of the 4d supergravity theory and these results can be found in the ancillary file. For the five supersymmetric backgrounds in Table~\ref{tab:prl-stable-vacua} these spectra can be organized into supersymmetric multiplets. Here we present the explicit form of these multiplets. Despite the fact that there is no conformal invariance each mode in the multiplet can be assigned a scaling weight (or dimension) and the scaling dimensions of members of the same supermultiplet differ by multiples of $1/2$. This is compatible with assigning a scaling weight of $1/2$ to the supercharges as familiar from 3d superconformal theories. We are not aware of a systematic discussion of supermultiplets for the type of scale-covariant supersymmetric QFTs of interest here but from the results summarized below it is clear that there is a close parallel with the structure of superconformal multiplets familiar from 3d SCFT, see \cite{Cordova:2016emh} for a summary.

We stress that the supermultiplets below contain only the fields retained by four-dimensional maximal supergravity. There is an infinite tower of higher KK modes arising from the 10d type IIA supergravity which we have not analyzed. We refer to~\cite{Bobev:2026xxx} for further discussion and more details on the holographic dictionary for scale-covariant backgrounds.

In the discussion below we use $\Delta$ to denote the lowest scaling weight in a given supermultiplet, and $S_\Delta$, $\chi_\Delta$, $A_\Delta$, $\psi_\Delta$ and $h_\Delta$ denote components of spin $0$, $\tfrac12$, $1$, $\tfrac32$ and $2$, respectively. The preserved supercharges raise the scaling weight by $\tfrac12$. 

\subsection{The \texorpdfstring{$\mathcal N=8$}{N=8} background \texttt{Z666666}}

\noindent The maximally supersymmetric D2-brane background has $\eta=1/3$ and unbroken $\mathrm{SO}(7)$ symmetry. The 4d supergravity fields organize into a single $\mathcal T_8$ multiplet dual to the $\mathcal N=8$ stress-tensor multiplet of 3d MSYM theory. The supercharges transform in the spinor $\mathbf8$ of the $\mathrm{SO}(7)$ R-symmetry. The physical components of the multiplet are
\begin{equation}
	\mathcal T_8:
	\qquad \mathbf{27} \times S_{d+\eta-2} \longrightarrow \mathbf{48}\times \chi_{d+\eta-\frac32} \longrightarrow \mathbf{35} \times S_{d+\eta-1} + (\mathbf{21}\oplus \mathbf{7}) \times A_{d+\eta-1} \longrightarrow \mathbf{8}\times \psi_{d+\eta-\frac12} \longrightarrow h_{d+\eta}\,.
	\label{eq:supp-z0-multiplet}
\end{equation}
Some comments are in order. The singlet metric mode is dual to the trace of the stress-energy tensor, which is non-vanishing as a consequence of broken conformal invariance. Similarly, the gravitino fluctuations contain $\mathbf{8}$ spin-$\frac12$ components that couple to the gamma-trace of the supercurrents. The $\mathbf{7}$ vector modes are dual to non-conserved currents, whose divergences are in turn dual to the longitudinal modes of the massive vectors. By contrast, the $\mathbf{21}$ vector modes are massless and dual to the conserved $\mathrm{SO}(7)$ currents of the SYM theory. Note that, in the absence of conformal symmetry, a protected scaling weight does not by itself imply current conservation. This point will be discussed in more detail in \cite{Bobev:2026xxx}. The resulting spectrum contains $128$ bosonic and $128$ fermionic degrees of freedom.

\subsection{The \texorpdfstring{$\mathcal N=2$}{N=2} background \texttt{Z570472}}

\noindent For this background we have $\eta=(R-7)/6$ with $R=\sqrt{73}$. The unbroken global symmetry is $\mathrm{SU}(2)_F\times\mathrm{U}(1)_R$ where $\mathrm{U}(1)_R$ is the 3d $\mathcal{N}=2$ R-symmetry. The matter representations are denoted by $\mathbf d_q$ where  $\mathbf d$ is the dimension of the $\mathrm{SU}(2)_F$ representation and $q$ is the $\mathrm{U}(1)_R$ charge normalized such that the two supercharges have $q=\pm2$.  The eight gravitini of the supergravity theory decompose as
\begin{equation}
	\mathbf8=\mathbf1_{-2}\oplus\mathbf1_{+2}
	\oplus2\times\mathbf1_0\oplus\mathbf2_{-1}\oplus\mathbf2_{+1}\,,
\end{equation}
where the first two entries on the right hand side indicate the preserved supersymmetry generators. The associated stress-tensor multiplet in the dual QFT contains the $\mathrm{U}(1)_R$ current, the two supercurrents and the stress tensor.  The other three massless vectors belong to $\mathrm{SU}(2)_F$ flavor-current multiplets.  The component structure of these protected supermultiplets is as follows
\begin{align}
	\mathcal T_2 & : \quad A_{d+\eta-1} \longrightarrow 2\times\psi_{d+\eta-1/2} \longrightarrow h_{d+\eta}\,, \\
	\mathcal J_2 & : \quad S_{d+\eta-2} \longrightarrow 2\times\chi_{d+\eta-3/2} \longrightarrow (S\oplus A)_{d+\eta-1}\,.
	\label{eq:supp-n2-protected}
\end{align}
The remaining fields organize into massive-gravitino multiplets $\mathcal G_2$, chiral and antichiral multiplets $\mathcal B_2$ and $\overline{\mathcal B}_2$, and short and long massive-vector multiplets $\mathcal V_{2,\rm S}$ and $\mathcal V_{2,\rm L}$ with the following component structure:
\begin{align}
	\mathcal G_2(\Delta) & :\quad \chi_\Delta\longrightarrow S_{\Delta+1/2}\oplus2\times A_{\Delta+1/2} \longrightarrow 2\times\chi_{\Delta+1} \oplus\psi_{\Delta+1} \longrightarrow A_{\Delta+3/2}\,,\\
	\mathcal B_2(\Delta),\ \overline{\mathcal B}_2(\Delta) & :\quad S_\Delta \longrightarrow \chi_{\Delta+1/2} \longrightarrow S_{\Delta+1}\,, \\
	\mathcal V_{2,\rm S}(\Delta) & :\quad S_\Delta \longrightarrow 2\times\chi_{\Delta+1/2} \longrightarrow 2\times S_{\Delta+1}\oplus A_{\Delta+1} \longrightarrow \chi_{\Delta+3/2}\,, \\
	\mathcal V_{2,\rm L}(\Delta) & :\quad S_\Delta \longrightarrow 2\times\chi_{\Delta+1/2} \longrightarrow 3\times S_{\Delta+1} \oplus A_{\Delta+1} \longrightarrow 2 \times \chi_{\Delta+3/2} \longrightarrow S_{\Delta+2}\,.
	\label{eq:supp-n2-templates}
\end{align}
The full supermultiplet spectrum is summarized in Table~\ref{tab:supp-z9-multiplets}. The $\mathrm{SU}(2)_F$ representation of each field in a supermultiplet is the same while the ${ \rm U(1)}_R$ charge changes by $\pm2$ after an action by a supercharge. 

These supergravity multiplets couple to 3d $\mathcal N=2$ multiplets in the dual field theory. While $\mathcal T_2$ contains the modes dual to the $\mathrm{U}(1)_R$ current, supercurrents and stress tensor, those dual to $\gamma\cdot S$ and $T^\mu{}_\mu$ belong to the trace multiplet $\mathcal V_{2,\rm L}\left((R-1)/6\right)$. Despite this supergravity split, the dual operators belong to a single $\mathcal N=2$ S-multiplet, linked by its supercurrent conservation equations~\cite{Dumitrescu:2011iu}. For $\mathcal N=8$, enhanced supersymmetry unifies these modes already in supergravity, yielding the single multiplet in~\eqref{eq:supp-z0-multiplet}.

\begin{table}[htb]
	\caption{Supersymmetric multiplets for \texttt{Z570472} with $R=\sqrt{73}$.
		The representation labels are specified in the text; the two
		neutral massive-gravitino multiplets have the same scaling weight.}
	\label{tab:supp-z9-multiplets}
	\centering
	\renewcommand{\arraystretch}{1.4}
	\begin{ruledtabular}
		\begin{tabular}{ccc}
			Family                 & $\mathrm{SU}(2)_F\times\mathrm{U}(1)_R$ representation        & $\Delta$          \\
			\hline
			$\mathcal T_2$         & $\mathbf1_0$                                           & $(R+5)/6$ \\
			$\mathcal J_2$         & $\mathbf3_0$                                           & $(R-1)/6$ \\
			$\mathcal G_2$         & $\mathbf2_{-1}\oplus\mathbf2_{+1}$                     & $(41+R)/24$  \\
			$\mathcal G_2$         & $2\times\mathbf1_0$                                          & $(5R-17)/12$ \\
			$\mathcal B_2\oplus\overline{\mathcal B}_2$
			& $\mathbf3_{+2}\oplus\mathbf3_{-2}$                     & $(R-3)/4$    \\
			$\mathcal V_{2,\rm S}$ & $\mathbf2_{-1}\oplus\mathbf2_{+1}$                     & $(71-5R)/24$ \\
			$\mathcal V_{2,\rm L}$ & $\mathbf1_0$                                           & $(R-1)/6$    \\
			$\mathcal V_{2,\rm L}$ & $\mathbf1_0$
			& $\dfrac{R-1}{12}+\dfrac12\sqrt{\dfrac{1789-169R}{18}}$                \\[5pt]
		\end{tabular}
	\end{ruledtabular}
\end{table}

\subsection{The \texorpdfstring{$\mathcal N=1$}{N=1} backgrounds}

\noindent For the $\mathcal{N}=1$ backgrounds we find the following supersymmetric multiplets 
\begin{equation}
	\begin{aligned}
		\mathcal T_1 & : \quad \psi_{d+\eta-1/2} \longrightarrow h_{d+\eta}\,,\\ 
		\mathcal J_1 & : \quad \chi_{d+\eta-3/2} \longrightarrow A_{d+\eta-1}\,,\\
		\mathcal G_1(\Delta) & : \quad A_\Delta \longrightarrow (\chi\oplus\psi)_{\Delta+1/2} \longrightarrow A_{\Delta+1}\,,\\
		\mathcal V_1(\Delta) & : \quad \chi_\Delta \longrightarrow (S\oplus A)_{\Delta+1/2} \longrightarrow \chi_{\Delta+1}\,,\\
		\mathcal S_1(\Delta) & : \quad S_\Delta \longrightarrow \chi_{\Delta+1/2} \longrightarrow S_{\Delta+1}\,.
	\end{aligned}
	\label{eq:supp-n1-multiplets}
\end{equation}
Here $\mathcal T_1$ and $\mathcal J_1$ are the stress-tensor and flavor-current multiplets, while $\mathcal G_1$, $\mathcal V_1$ and $\mathcal S_1$ are the massive-gravitino, massive-vector and scalar multiplets. All three $\mathcal N=1$ spectra contain $128$ bosonic and $128$ fermionic degrees of freedom.

\medskip

\paragraph{The $\mathrm{SU}(3)$ background \texttt{Z597968}.}
The preserved supercharge is an $\mathrm{SU}(3)$ singlet and $\eta=(\sqrt{1569}-33)/24$. The graviton and one gravitino form $\mathcal T_1$, and the eight unbroken vectors form flavor-current multiplets in the adjoint of $\mathrm{SU}(3)$. The seven massive gravitini transform as $\mathbf1 \oplus \mathbf3 \oplus \overline{\mathbf3}$. Table~\ref{tab:supp-z4-multiplets} lists every multiplet, including its lowest scaling weight.  In total there are one stress-tensor, eight current, seven massive-gravitino, six massive-vector and twenty-two scalar multiplets. 
\medskip

\paragraph{The $\mathrm{U}(1)$ backgrounds \texttt{Z448357} and \texttt{Z394258}.}
Each background preserves one supercharge and has one Abelian flavor current. Both spectra contain the following multiplets:
\begin{equation}
	\mathcal T_1 \oplus \mathcal J_1 \oplus \bigoplus_{a=1}^{7}\mathcal G_1(\Delta_a) \oplus \bigoplus_{b=1}^{13}\mathcal V_1(\Delta_b) \oplus \bigoplus_{c=1}^{15}\mathcal S_1(\Delta_c)\,.
	\label{eq:supp-u1-multiplets}
\end{equation}
The two protected multiplets have the weights in Eq.~\eqref{eq:supp-n1-multiplets}, with $d+\eta=3.164959999154\ldots$ and $d+\eta=3.143392795482\ldots$, respectively. The remaining weights are listed separately in Table~\ref{tab:supp-u1-multiplets}. The numerical values for $\eta$ and the scaling weights are algebraic numbers found as solutions of high order polynomial equations. 

Similar to the $\mathcal{N}=2$ case above, the multiplet $\mathcal T_1$ contains the modes dual to the supercurrent and stress tensor, while the singlet scalar multiplet $\mathcal S_1(d+\eta-1)$, present in all three spectra, contains those dual to $\gamma\cdot S$ and $T^\mu{}_\mu$. In the dual QFT, these operators belong to a single 3d $\mathcal N=1$ supercurrent multiplet, linked by its supercurrent conservation equations~\cite{Drukker:2017dgn}.

\makeatletter
\setlength{\@fpsep}{10pt}
\makeatother

\begin{table}[htb]
	\caption{Supersymmetric multiplets of \texttt{Z597968}.  Each row represents a multiplet transforming in the indicated $\mathrm{SU}(3)$ representation.  The scaling weight $\Delta$ is that of the first component in Eq.~\eqref{eq:supp-n1-multiplets}; numerical entries are rounded to nine decimal places.}
	\label{tab:supp-z4-multiplets}
	\centering
	\begin{ruledtabular}
		\begin{tabular}{ccc}
			Family         & $\mathrm{SU}(3)$ representation     & $\Delta$ \\
			\hline
			$\mathcal T_1$ & $\mathbf1$                          & $d+\eta-\tfrac12$ \\
			$\mathcal J_1$ & $\mathbf8$                          & $d+\eta-\tfrac32$ \\
			$\mathcal G_1$ & $\mathbf3\oplus\overline{\mathbf3}$ & $2.478148038$ \\
			$\mathcal G_1$ & $\mathbf1$                          & $2.630568867$ \\
			$\mathcal V_1$ & $\mathbf3\oplus\overline{\mathbf3}$ & $1.757511367$ \\
			$\mathcal S_1$ & $\mathbf6\oplus\overline{\mathbf6}$ & $1.514632939$ \\
			$\mathcal S_1$ & $\mathbf8$                          & $1.712492541$ \\
			$\mathcal S_1$ & \multicolumn{1}{c}{$\mathbf1$}      & $d+\eta-1$ \\
			$\mathcal S_1$ & $\mathbf1$                          & $3.735723668$\\
		\end{tabular}
	\end{ruledtabular}
\end{table}

\begin{table}[htb]
	\caption{Lowest scaling weights of the long multiplets at \texttt{Z448357} and \texttt{Z394258}.
		A superscript $(n)$ denotes multiplicity; entries without a superscript have multiplicity one.  Each column contains seven $\mathcal G_1$, thirteen $\mathcal V_1$ and fifteen $\mathcal S_1$ multiplets, in addition to the stress-tensor and current multiplets.}
	\label{tab:supp-u1-multiplets}
	\centering\small
	\renewcommand{\arraystretch}{1.25}
	\begin{ruledtabular}
		\begin{tabular}{ccc}
			Family & $\Delta$ at \texttt{Z448357} & $\Delta$ at \texttt{Z394258} \\
			\hline
			$\mathcal G_1$ 
			& 
			$\begin{gathered}
				2.670678959^{(2)},\ 2.701089927^{(2)},\ 2.841094265,\\
				2.924606628,\ 3.127378802
			\end{gathered}$ 
			& 
			$\begin{gathered}
				2.804365706^{(2)},\ 2.855971816,\ 2.899224428,\\
				2.924081605,\ 2.969512963^{(2)}
			\end{gathered}$             
			\\[8pt]
			\hline
			$\mathcal V_1$ 
			& 
			$\begin{gathered}
				1.511170048,\ 1.651421501^{(2)},\ 1.760186398^{(2)},\\
				2.392774229,\ 2.432590356,\ 2.935659704^{(2)},\\
				3.717819807^{(2)},\ 3.747800275^{(2)}
			\end{gathered}$ 
			& 
			$\begin{gathered}
				1.544363090^{(2)},\ 1.959390941^{(2)},\ 2.048511711,\\
				2.075902877^{(2)},\ 3.749539408^{(2)},\ 3.773947017^{(2)},\\
				3.851402132,\ 3.912416584
			\end{gathered}$ 
			\\[8pt]
			\hline
			$\mathcal S_1$ 
			& 
			$\begin{gathered}
				1.404940388,\ 1.476473199,\ 1.551095823^{(2)},\\
				1.560056581^{(2)},\ 1.698359529,\ 2.135884712^{(2)},\\
				2.164959999,\ 3.670903691,\ 3.674118827,\\
				3.864995126,\ 4.224735994^{(2)}
			\end{gathered}$ 
			& 
			$\begin{gathered}
				1.333808038,\ 1.425712751,\ 1.615210934^{(2)},\\
				1.730474034^{(2)},\ 2.143392795,\ 2.819637067^{(2)},\\
				3.087966834^{(2)},\ 3.193200215,\ 3.822733709,\\
				4.824289749,\ 4.842402409
			\end{gathered}$       
			\\[3pt]
		\end{tabular}
	\end{ruledtabular}
\end{table}

\end{document}